# Low half-wave-voltage, ultra-high bandwidth thin-film LiNbO₃ modulator based on hybrid waveguide and periodic capacitively loaded electrodes

XUECHENG LIU, BING XIONG,* CHANGZHENG SUN, JIAN WANG, ZHIBIAO HAO, LAI WANG, YANJUN HAN, HONGTAO LI, JIADONG YU, YI LUO

*Beijing National Research Center for Information Science and Technology (BNRist), Department of Electronic Engineering, Tsinghua University, Beijing100084, China*
*\*Corresponding author: bxiong@tsinghua.edu.cn*

**A novel thin-film LiNbO₃ (TFLN) electro-optic modulator is proposed and demonstrated. LiNbO₃-silica hybrid waveguide is adopted to maintain low optical loss for an electrode spacing as narrow as 3 μm, resulting in a record low half-wave-voltage length product of only 1.7 V·cm. Capacitively loaded traveling-wave electrodes (CL-TWEs) are employed to reduce the microwave loss, while quartz substrate is used in place of silicon substrate to achieve velocity matching. The fabricated TFLN modulator with a 5-mm-long modulation region exhibits a half-wave-voltage of 3.4 V and merely 1.3 dB roll-off in electro-optic response up to 67 GHz, and a 3-dB modulation bandwidth over 110 GHz is predicted.**

High-speed electro-optic modulators are key devices for high-capacity fiber-optic communications [1] and microwave-photonic links [2]. Lithium niobate (LiNbO₃, LN) has been the preferred material for electro-optic modulators, thanks to its high electro-optic coefficient ($r_{33}$ = 31 pm/V), wide transmission window (340 ~ 4600 nm) and low optical loss at telecom wavelengths. However, traditional LN waveguides based on titanium in-diffusion or proton-exchange exhibit a low refractive index contrast below 0.1 [3,4], which leads to large optical mode size, high bending loss, low power efficiency and difficulty in integration. Photonic platforms based on silicon [5,6], polymers [7,8], and III-V compound semiconductors [9,10] have also been proposed, but none of them can secure wide modulation bandwidth, low drive voltage and low insertion loss simultaneously.

Recently, thin-film lithium niobate (TFLN) fabricated by crystal ion slicing and wafer bonding [11] has proved particularly attractive for realizing compact integrated devices. The high refractive index contrast of the TFLN ridge waveguide allows enhanced modulation efficiency. So far, TFLN Mach-Zehnder modulators (MZMs) with improved performances over legacy LN modulators have been demonstrated [12-16], including low half-wave-voltage, large modulation bandwidth and small footprint.

For many applications, it is desirable to further reduce the half-wave-voltage length product $V_\pi L$ and extend the modulation bandwidth of TFLN modulators. The half-wave-voltage length product for previously reported devices is mainly limited by the electrode spacing, which is mostly beyond 5 μm to avoid excessive optical absorption loss [12-16]. On the other hand, the modulation bandwidth depends critically on the microwave loss of the traveling wave electrodes provided that impedance matching and velocity matching are satisfied [17].

To break these performance limitations, we propose a TFLN modulator based on LiNbO₃-silica hybrid waveguide, which allows an electrode gap as narrow as 3 μm for enhanced electric field loading efficiency. Meanwhile, capacitively-loaded traveling-wave electrodes (CL-TWEs) are adopted to reduce the microwave loss. Furthermore, to overcome the slow wave effect of the CL-TWEs, quartz substrate with low dielectric constant is employed to implement velocity matching between the microwave and the optical signals [18]. Compared with TFLN modulators on silicon substrate [19], TFLN modulators formed on quartz substrate are found to exhibit significantly improved high-frequency response.

Figure 1(a) depicts the ridge waveguide structure commonly adopted in a TFLN modulator. The refractive index variation in the X-cut LN waveguide is given by [20]

$$\Delta n_{eff} = \frac{1}{2}\Gamma\gamma_{33}n_o^3\frac{V}{g} \quad (1)$$

where $\Gamma$ is the electro-optic overlap integral, $\gamma_{33}$ is the electro-optic coefficient, $n_o$ is the optical refractive index, while $V$ and $g$ are the modulation voltage and the gap between the electrodes, respectively. Reducing the electrode spacing $g$ helps increase

Δ$n_{eff}$, but also leads to enhanced optical loss at the metal electrodes. To avoid excessive optical loss, the electrode spacing in most TFLN modulators is beyond 5 μm, thus limiting the modulation efficiency. In this work, a LiNbO$_3$-silica hybrid waveguide is adopted, as shown in Fig. 1(b), which is formed by covering the partially etched LiNbO$_3$ ridge waveguide with a thin silica buffer layer. Our simulations reveal that the silica buffer layer helps suppress the surface plasmon polariton (SPP) mode at the LiNbO$_3$/metal interface, thus reducing the optical absorption loss at the metal electrodes dramatically (See Supplementary material). As illustrated in Fig. 1(c), for a 1-μm-wide and 300-nm-thick partially etched ridge waveguide formed on a 600-nm-thick TFLN, the introduction of a 100-nm-thick silica buffer layer helps reduce the optical loss by more than two orders of magnitude. As a result, an optical loss less than 0.1 dB/cm can be maintained for an electrode gap as narrow as 3 μm. As illustrated in Figs. 1(d) and 1(e), the 3-μm-spaced electrodes formed on the hybrid waveguide leads to an RF electric field enhancement of more than 40% in the LN ridge waveguide, thus reducing the half-wave-voltage of the modulator effectively.

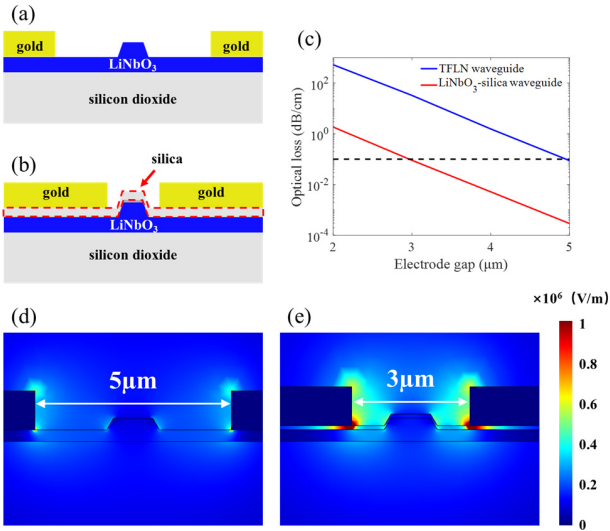

Fig. 1. (a) Conventional TFLN waveguide with wide-gap electrodes. (b) LiNbO$_3$-silica hybrid waveguide with narrow-gap electrodes. (c) Optical absorption loss of optical waveguides with/without the silica buffer layer. RF modes in (d) TFLN waveguide with wide-gap electrodes and (e) LiNbO$_3$-silica hybrid waveguide with narrow-gap electrodes.

The thickness of the silica buffer layer is optimized for a trade-off between a small electrode spacing and a large electro-optic overlap factor. Assuming an upper-limit of 0.1 dB/cm for optical absorption loss, the minimum allowable electrode spacing and the corresponding electro-optic overlap factor calculated by finite element method (FEM) are plotted in Fig. 2(a). It is evident that the minimum electrode spacing reduces rapidly as the silica buffer layer thickness increases, whereas the electro-optic overlap factor shows only a moderate reduction. The refractive index variation in the LiNbO$_3$-silica hybrid waveguide under 1 V drive voltage is shown in Fig. 2(b). A 100-nm-thick silica buffer layer with an electrode spacing of 3 μm is found to provide the optimum modulation efficiency.

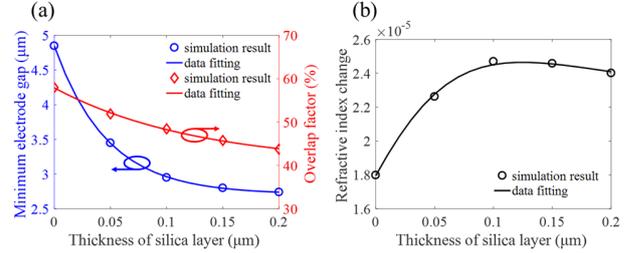

Fig. 2. (a) Minimum allowable electrode spacing and electro-optic overlap factor for different silica buffer layer thicknesses. (b) Refractive index variation of the hybrid waveguide with 1 V drive voltage. A 100-nm-thick silica layer is chosen to obtain the maximum refractive index change.

The key to extending the modulation bandwidth lies in reducing the microwave loss while implementing velocity and impedance matching [17]. The microwave loss of the modulator comes from the traveling-wave electrodes as well as the substrate absorption. A narrow electrode spacing facilitates electro-optic interaction, but also means a narrow signal electrode for impedance matching [21], which tends to increase the microwave transmission loss. To reduce the microwave loss while maintaining the narrow electrode gap required for low drive voltage, we adopt traveling-wave electrodes with periodically loaded T-rails, as shown in Fig. 3(a). Narrow-gap T-rails can realize high-efficiency electric field loading, while a wide signal electrode helps ensure low-loss microwave transmission. Such CL-TWEs were previously employed in III-V semiconductor electro-optic modulators [11], and have recently been proposed for TFLN modulators [18,22]. As long as the T-rails are much smaller than the microwave wavelength, they act as capacitive loadings and contribute little to the microwave transmission loss. Consequently, both wide modulation bandwidth and low half-wave-voltage can be realized simultaneously, thus breaking the voltage-bandwidth limitation of TFLN modulators. As the CL-TWEs form a slow-wave structure, TFLN bonded to low dielectric constant quartz substrate is adopted instead of TFLN on silicon [12]. The quartz substrate not only helps implement velocity matching by counteracting the slow wave effect of the CL-TWEs, but also reduces the microwave loss (See Supplementary material).

The characteristic impedance and the refractive index of the traveling-wave electrodes can be expressed as

$$Z_m = \sqrt{\frac{L_0}{C_0}} \quad (2)$$

and

$$n_m = \frac{c}{v_m} = c\sqrt{L_0 C_0} \quad (3)$$

where $C_0$ and $L_0$ are the capacitance and inductance per unit length, respectively. Since the capacitance of the CL-TWEs mainly comes from the T-rails, whereas the inductance mainly depends on the width of the central signal electrode and the spacing between the unloaded electrodes, impedance and velocity matching can be implemented by adjusting the dimensions of the loaded/unloaded electrodes [23]. According to our FEM simulations shown in Fig. 3(b), a duty cycle of 90% is required for the T-rails spaced 3 μm apart. Figure 3(c) plots the microwave loss of the CL-TWEs as a function of the central signal electrode width. The microwave loss consists of two parts [24]: the microwave loss of the unloaded electrodes dominates for narrow signal electrode, whereas the additional loss caused by T-rails becomes significant as the signal electrode widens. To ensure low microwave loss, the signal electrode width and unloaded electrode spacing are chosen to be 50 μm and 15 μm, respectively. In addition, the period of the T-rails is set as 50 μm, so as to ensure a large cut-off frequency in this periodic electrode structure [10].

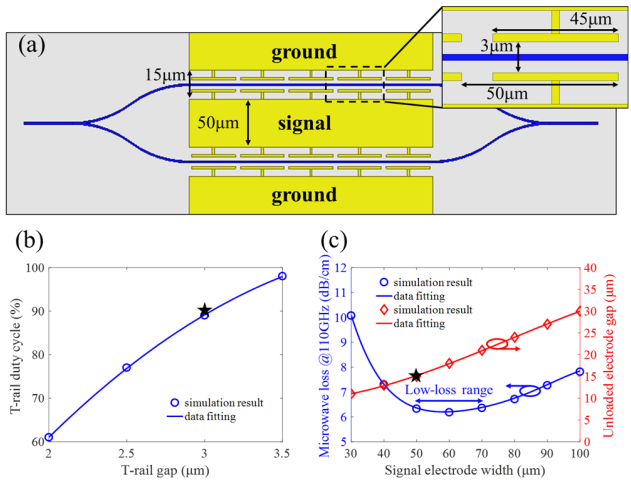

Fig. 3. (a) Top view of the CL-TWEs. (b) Duty cycle of T-rails for different T-rail gaps under capacitance matching condition. (c) Variation of microwave loss with the width of unloaded signal electrode under inductance matching condition. The star in (b) and (c) indicates the designed value.

The modulator is fabricated with a wafer of TFLN on quartz provided by NanoLN, which includes a 2-μm-thick $SiO_2$ bonding layer between the 600-nm-thick X-cut TFLN and the 500-μm-thick quartz substrate. The optical waveguide is patterned by electron beam lithography (EBL) with hydrogensilsesquioxane (HSQ), and transferred to the TFLN by argon based reactive ion etching (RIE). The TFLN is partially etched by 300 nm, and then covered with a 100-nm-thick $SiO_2$ layer by plasma enhanced chemical vapor deposition (PECVD). A two-step fabrication process is employed for the CL-TWEs: First a lift-off process with polymethylmethacrylate (PMMA) exposed by EBL is employed to form T-rails with high position accuracy. Next, the main electrode patterns are defined by contact UV lithography. The main electrodes are then thickened to 1.4 μm by electroplating to reduce the microwave loss. The 3D schematic of the TFLN modulator with CL-TWEs is shown in Fig. 4(a). The scanning electron microscope (SEM) image of the T-rail electrodes is shown as inset. To facilitate characterization of the modulation response at high frequencies with microwave probes, the CL-TWEs are converted to standard CPW electrodes outside the modulation region. The CPW electrodes with 90° bend are formed on 1-μm-thick benzocyclobutene (BCB) cladding layer to reduce the optical absorption (See Supplementary material).

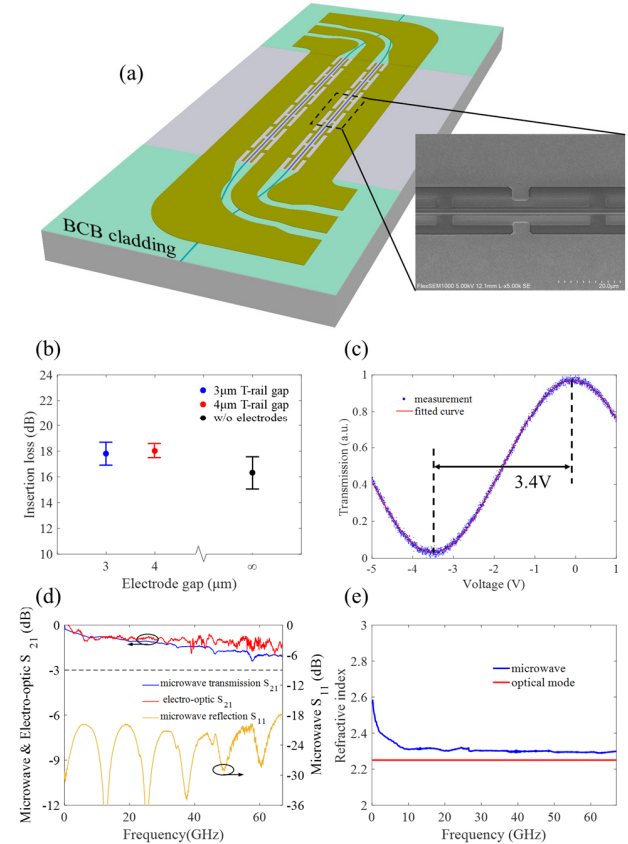

Fig. 4. (a) 3D schematic of the demonstrated modulator. The unloaded electrodes have 50 μm signal electrode width, 15 μm electrode spacing and bended tapers to match with the microwave probes. The inset shows the SEM image of the 3-μm spaced T-rails with 50 μm period and 90% duty cycle. (b) Insertion loss of 5-mm-long modulators with different electrode gaps. (c) Normalized optical transmissions as a function of modulation voltage. (d) Microwave transmission $S_{21}$ and reflection $S_{11}$ of the traveling-wave electrodes as well as the electro-optic response of the TFLN modulator up to 67 GHz. (e) The extracted microwave refractive index, which shows excellent matching with the group index of the optical mode ($n_g$ ~2.25).

The performance of the modulator is tested by end-butt coupling with two tapered fibers, and a polarization controller is used to ensure TE polarized incident light at 1550 nm. In Fig. 4(b), we plot the insertion loss of 5-mm-long modulators with T-rail gaps of 3 and 4 μm. Compared with the device without electrodes, it can be concluded that the excessive optical loss due to narrow electrode gaps is effectively suppressed by the silica buffer layer, in agreement with the estimation shown in

Fig. 1(c). The total insertion loss is measured to be 17 dB for a 5-mm-long modulator, which mainly comes from the coupling loss, as the mode field mismatch between the fiber and the optical waveguide has not been optimized. The coupling efficiency can be significantly improved by employing an inverse taper [25]. For half-wave-voltage characterization, a 100 kHz triangular wave signal is applied to the modulator, and the modulated signal is captured with a photodetector connected to an oscilloscope, as shown in Fig. 4(c). The extracted half-wave-voltage for the 5-mm-long modulator is 3.4 V, together with an extinction ratio beyond 17 dB. The half-wave-voltage length product $V_\pi L$ is as low as 1.7 V·cm, in excellent agreement with our simulations. To the best of our knowledge, this is the lowest $V_\pi L$ reported for TFLN modulators, which mainly benefits from the narrow electrode gap of 3 μm.

A frequency response test system with a bandwidth up to 67 GHz is used to characterize the electro-optic response of the modulator. First, the microwave transmission $S_{21}$ and reflection $S_{11}$ of the traveling-wave electrodes are measured by an Agilent N5227A vector network analyzer (VNA) with ground–signal–ground (GSG) microwave probes, as shown in Fig. 4(d). The microwave transmission shows a roll-off less than 2 dB up to 67 GHz, while the electrical reflection remains below −18 dB over the entire testing frequency range, indicating good impedance matching. The extracted refractive index for microwave signal is close to the optical group index ($n_g \sim 2.25$), as shown in Fig. 4(e), implying excellent velocity matching. To further characterize the electro-optic frequency response, the modulation signal from the VNA is fed to the modulator through a microwave probe, and the modulated signal is fed back to the VNA via a high-speed photodetector (FINISAR XPDV3120R), while another microwave probe is used to provide a 50 Ω impedance termination. The electro-optic response after calibration is also plotted in Fig. 4(d). Thanks to excellent impedance and velocity matching, the modulator exhibits a smooth electro-optic frequency response, and only a 1.3 dB roll off is recorded at 67 GHz, which is limited by the bandwidth of the test system. By adopting a VNA with a bandwidth up to 110 GHz (Keysight N5290A), we have verified that the microwave transmission of the traveling-wave electrodes exhibits a 6-dB bandwidth over 110 GHz (See Supplementary material). Based on the extracted microwave refractive index, a 3-dB modulation bandwidth over 110 GHz is predicted.

In this work, we have proposed a TFLN modulator structure capable of both wide modulation bandwidth and low half-wave-voltage length product. Modulators equipped with 3 μm electrode spacing CL-TWEs are fabricated on TFLN wafer bonded to quartz substrate. A half-wave-voltage length product as low as 1.7 V·cm has been demonstrated with a LiNbO$_3$-silica hybrid waveguide. The electro-optic response of a device with 5 mm modulation length shows only 1.3 dB roll-off up to 67 GHz, and a modulation bandwidth exceeding 110 GHz is expected.

**Funding.** This work was supported in part by National Key R&D Program of China (2018YFB2201701); National Natural Science Foundation of China (61975093, 61927811, 61991443, 61822404, 61974080, 61904093, and 61875104); Key Lab Program of BNRist (BNR2019ZS01005); China Postdoctoral Science Foundation (2019T120090) and Collaborative Innovation Centre of Solid-State Lighting and Energy-Saving Electronics.

**Acknowledgment.** The authors thank Prof. Xinlun Cai of Sun Yet-sen University for help in high frequency electro-optic response measurement.

# Low half-wave-voltage, ultra-high bandwidth thin-film LiNbO$_3$ modulator based on hybrid waveguide and periodic capacitively loaded electrodes: supplemental document

This document provides supplementary information to "Low half-wave-voltage, ultra-high bandwidth thin-film LiNbO$_3$ modulator based on hybrid waveguide and periodic capacitively loaded electrodes."

## 1. Analysis of optical absorption loss caused by metal electrodes

The optical waveguide fabricated on thin film lithium niobate (TFLN) by dry etching has a mode field diameter an order of magnitude smaller than that of the waveguide formed on bulk LiNbO$_3$ [S1]. However, the electrode gap of the X-cut TFLN modulators is not decreased in the same proportion, as most reported value is beyond 5 μm [S2-S6]. This is mainly attributed to the fact that the absorption loss due to metal electrodes increases exponentially as the electrode gap decreases, as shown in Fig. S1(a). For narrow-gap electrodes, the optical field is strongly coupled to the surface-plasmon-polariton (SPP) mode excited at the metal-LiNbO$_3$ interface, which is distributed along the surface of the metal electrodes, as shown in Fig. S1(c). To better display the SPP mode, a logarithmic color scale is adopted. The excitation of SPP mode would lead to excessive optical loss for a TFLN waveguide with narrow-gap electrodes. For LiNbO$_3$-silica hybrid waveguides, it can be seen in Fig. S1(d) that a metal-silica-LiNbO$_3$ interface inhibits the excitation of SPP mode. As a result, optical absorption loss can be reduced by more than two orders of magnitude at the same electrode gap.

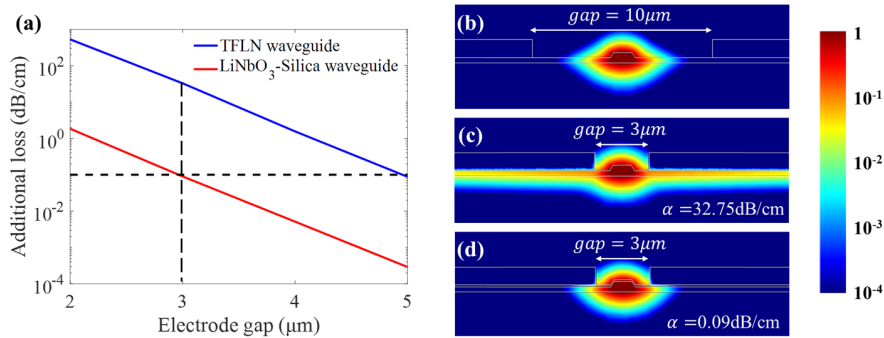

Fig. S1. (a) The suppression effect of LiNbO$_3$-Silica hybrid waveguide on optical absorption loss. (b) Optical mode field in TFLN waveguide with wide-gap electrodes. (b) Optical mode field in TFLN waveguide with narrow-gap electrodes. (b) Optical mode field in LiNbO$_3$-silica hybrid waveguide with narrow-gap electrodes. (All in logarithmic scale)

In the proposed TFLN modulator, two 90° bend coplanar waveguide (CPW) electrodes are formed at the end of the capacitively-loaded traveling-wave electrodes (CL-TWEs) for probe test. To minimize the optical loss due to the CPW tapers on top of the TFLN waveguides, electrodes, photosensitive benzocyclobutene (BCB) is formed between the optical waveguide and bended electrodes. The BCB layer in the modulation region is removed by photolithography. According to our simulations shown in Fig. S2(a), the optical absorption loss is negligible for a BCB layer thickness beyond 1 μm. The optical modes at different regions along the TFLN waveguide are shown in Figs. S2(b) and S2(c), respectively.

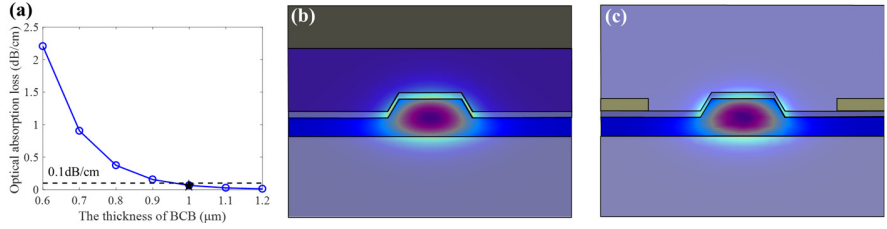

Fig. S2. (a) Optical absorption loss due to bended CPW electrodes varies with BCB layer thickness. The black star represents our designed value. (b) Optical mode field with BCB cladding and bended CPW electrodes. (c) Optical mode field in the modulation region.

## 2. Optimization of travelling-wave electrodes

The microwave transmission $S_{21}$ and reflection $S_{11}$ of the capacitively-loaded traveling-wave electrodes (CL-TWEs) are simulated by finite element method (FEM). The T-rail gap is taken to be 3 μm, while the duty cycle of T-rails has to be optimized. As the T-rails mainly serve as capacitive loading, the duty cycle mainly affects the capacitance of the CL-TWEs. On the other hand, the inductance of the CL-TWEs is almost identical to the unloaded value, i.e. the signal electrode width and the unloaded electrode spacing determine the inductance of the CL-TWEs.

The microwave loss of the CL-TWEs mainly consists of two parts [S7]

$$\alpha = \frac{R_0}{2Z_0} + N\left(\frac{I_T}{I}\right)^2 \frac{R_T}{2Z_0} \tag{S1}$$

where $R_0$ and $R_T$ are the surface resistances of the unloaded electrodes and the T-rails, respectively. $I$ and $I_T$ are the axial current along the unloaded electrodes and the T-rails, respectively. $Z_0$ is the characteristic impedance of the CL-TWEs, while $N$ is the number of the T-rails. The first term corresponds to the microwave loss of the unloaded electrodes, while the second term is the additional loss introduced by the T-rails. As the signal electrode widens, the surface resistance $R_0$ and the microwave loss of the unloaded electrode decreases. However, the T-rail resistance $R_T$ for a wide unloaded electrode gap would increase. We determine the low-loss range by a parameter scanning, and the signal electrode width and the unloaded electrode spacing are chosen to be 50 μm and 15 μm, respectively.

Another important issue for the design of the CL-TWEs is the cut-off frequency of the periodic structures, which can be estimated by

$$f_{cut-off} = \frac{c}{2n_m L_p} \tag{S2}$$

where $n_m$ is the refractive index of microwave signal, while $L_p$ is the period of T-rails. The influence of cut-off frequency on the refractive index and loss of the CL-TWEs are shown in Fig. S4. It is seen that both the microwave refractive index and microwave loss increases dramatically around the cut-off frequency. A relatively short T-rail period of 50 μm is chosen for our TFLN modulator, so that the cut-off frequency is beyond 800 GHz. The corresponding T-rail length is 45 μm at 90% duty cycle. The microwave transmission and reflection of the CL-TWEs with optimized parameters shown in Table S1 are plotted in Fig. 3(b). A 6-dB bandwidth over 200 GHz is predicted for the 5-mm-long CL-TWEs.

**Table S1. Parameters of the CL-TWEs**

| parameters | $L_p$ | $L_{act}$ | $L_T$ | $W_s$ | $W_T$ | $G_1$ | $G_2$ |
|---|---|---|---|---|---|---|---|
| value | 50μm | 45μm | 5μm | 50μm | 3μm | 3μm | 15μm |

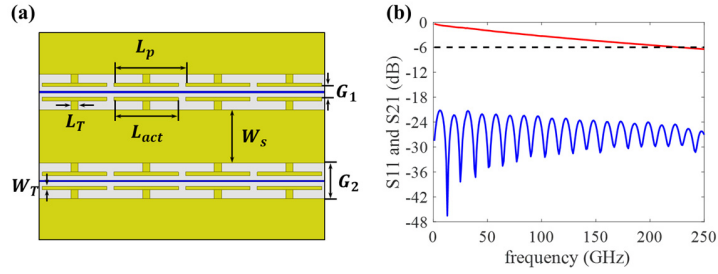

Fig. S3. (a) Structure of the CL-TWEs. (b) Simulated $S_{11}$ and $S_{21}$ of 5-mm-long CL-TWEs.

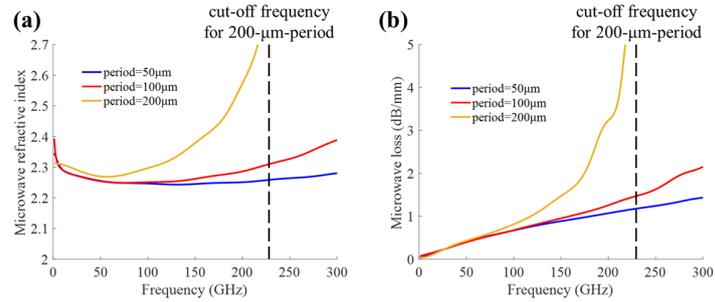

Fig. S4. The influence of cut-off frequency on (a) microwave refractive index and (b) microwave loss.

## 3. Electro-optic response characterization

For half-wave-voltage characterization, a 100 kHz triangular voltage signal is applied to the modulator, as the slow effect of LiNbO$_3$ such as electrical relaxation effect may result in measurement errors [S8]. The half-wave-voltage length product is extracted as 1.7 V·cm for 5 mm modulation length, while a more accurate extinction ratio of 17 dB is obtained by DC voltage scanning.

For high frequency electro-optic response measurement, a 90° bend CPW feedline is designed to facilitate microwave probe testing. Microwave transmission $S_{21}$ and reflection $S_{11}$ of the traveling-wave electrodes from 100 MHz to 110 GHz are measured with a Keysight N5290A vector network analyzer (VNA), as shown in Fig. S5(b). Some resonances occur in the high frequency range. Such resonances are also found in the simulated microwave S parameters of the traveling-wave electrodes with bended CPW feedlines, as shown in Fig. S5(a). They are attributed to mode mismatch between the CPW waveguide taper and the CL-TWEs, and can be avoided by further optimization in our future work. Nevertheless, the microwave transmission $S_{21}$ shows only a 3.7 dB roll-off at 104 GHz, and the 6-dB bandwidth exceeds the testing range of the VNA, i.e. 110 GHz.

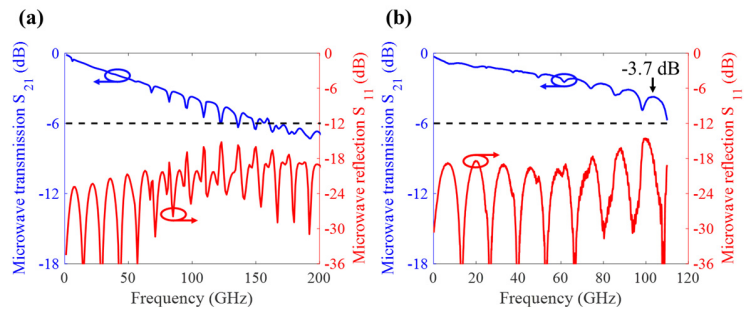

Fig. S5. (a) Simulated and (b) measured microwave S parameters of the CL-TWEs with bended CPW feedlines.

CL-TWEs with the same structure are also fabricated on TFLN with silicon substrates. As shown in Fig. S6(a), the microwave transmission $S_{21}$ shows a rapidly drop-off at low frequencies, which is attributed to additional absorption due to interfacial charges between the silicon substrate and the silicon dioxide layer [S9]. According to Fig. S6(b), where the microwave loss is plotted against $\sqrt{f}$, it can be concluded that the microwave loss of the CL-TWEs on quartz substrate mainly comes from ohmic loss [S10]. From this point of view, silicon is not an ideal substrate material for high-speed electro-optic modulators. Furthermore, the microwave refractive index is greatly reduced by the quartz substrate and well matched to the optical group index, as shown in Fig. S6(c). All of these prove the importance of using quartz substrate for a wide modulation bandwidth.

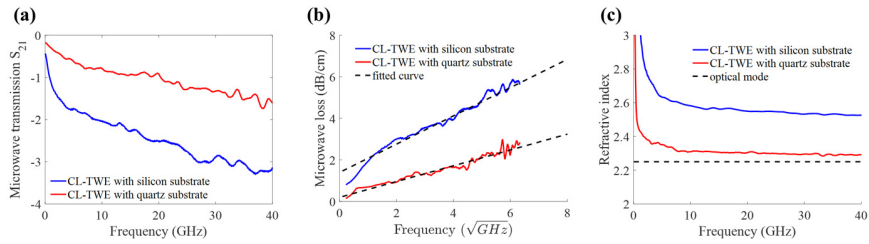

Fig. S6. (a) Microwave transmission ($S_{21}$) of CL-TWE with different substrates. (b) Relationship between microwave loss of CL-TWE and $\sqrt{f}$ with different substrates. (c) Microwave refractive index of CL-TWE with different substrates.

The electro-optic response shown in the main text is tested up to 67 GHz, which is limited by the bandwidth of the measurement system. As the electro-optic response curve only shows a 1.3 dB drop-off up to 67 GHz, a much wider 3-dB bandwidth is expected for the TFLN modulator. The electro-optic response up to 110 GHz is calculated from the microwave S-parameters by using the transmission line model [S11], as plotted in Fig. S7. It shows a good agreement within the measured frequency range. A 3-dB electro-optic bandwidth exceeding 110 GHz is predicted for our TFLN modulator based on CL-TWEs and quartz substrate.

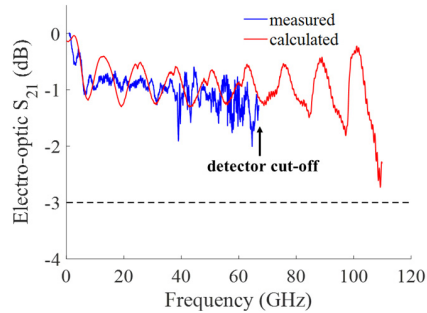

Fig. S7. Measured and calculated modulation response.